\documentclass[]{spie}  
\usepackage[]{graphicx}

\title{Precision velocimetry planet hunting with PARAS: \\Current performance and lessons to inform future \\extreme precision radial velocity instruments} 

\author{Arpita Roy\supit{a,b}, Abhijit Chakraborty\supit{c}, Suvrath Mahadevan\supit{a,b}, Priyanka Chaturvedi\supit{c}, Neelam J.S.S.V. Prasad\supit{c}, Vishal Shah\supit{c}, F.M. Pathan\supit{c}, and B. G. Anandarao\supit{c}
\skiplinehalf
{\normalsize 
\supit{a}Department of Astronomy \& Astrophysics, The Pennsylvania State University, University Park, 16802, USA. \\
\supit{b}Center for Exoplanets \& Habitable Worlds, The Pennsylvania State University, University Park, PA 16802. \\
\supit{c}Astronomy \& Astrophysics Division, Physical Research Laboratory, Ahmedabad 380009, India.
}}

\authorinfo{E-mail: arpita@psu.edu, Telephone: 1 814 863 5565}

\begin{document} 
  \maketitle 

\begin{abstract}
The PRL Advanced Radial-velocity Abu-sky Search (PARAS) instrument is a fiber-fed stabilized high-resolution cross-dispersed echelle spectrograph, located on the 1.2 m telescope in Mt. Abu India. Designed for exoplanet detection, PARAS is capable of single-shot spectral coverage of 3800 - 9600 \AA, and currently achieving radial velocity (RV) precisions approaching $\sim$ 1 m s$^{-1}$ over several months using simultaneous ThAr calibration. As such, it is one of the few dedicated stabilized fiber-fed spectrographs on small (1-2 m) telescopes that are able to fill an important niche in RV follow-up and stellar characterization. The success of ground-based RV surveys is motivating the push into extreme precisions, with goals of $\sim$ 10 cm s$^{-1}$ in the optical and $< $1 m s$^{-1}$ in the near-infrared (NIR). Lessons from existing instruments like PARAS are invaluable in informing hardware design, providing pipeline prototypes, and guiding scientific surveys. Here we present our current precision estimates of PARAS based on observations of bright RV standard stars, and describe the evolution of the data reduction and RV analysis pipeline as instrument characterization progresses and we gather longer baselines of data. Secondly, we discuss how our experience with PARAS is a critical component in the development of future cutting edge instruments like (1) the Habitable Zone Planet Finder (HPF), a near-infrared spectrograph optimized to look for planets around M dwarfs, scheduled to be commissioned on the Hobby Eberly Telescope in 2017, and (2) the NEID optical spectrograph, designed in response to the NN-EXPLORE call for an extreme precision Doppler spectrometer (EPDS) for the WIYN telescope. In anticipation of instruments like TESS and GAIA, the ground-based RV support system is being reinforced.  We emphasize that instruments like PARAS will play an intrinsic role in providing both complementary follow-up and battlefront experience for these next generation of precision velocimeters.
\end{abstract}



\section{INTRODUCTION}
\label{sec:intro}  
As we await the bounty of sources expected from space-based missions like TESS and GAIA, it is imperative that we tool up for the efficient and widespread ground-based follow-up of exoplanet candidates that will be required in complement. While confirmation of the most promising, and likely Earth analogous, candidates will be the prerogative of next-generation instruments like NEID and ESPRESSO, high-precision instruments on small (1-2~m) telescopes will be tasked with providing longer baseline observations on the rest. These instruments fill a niche that would be prohibitively expensive to fulfill with larger facilities, providing both high quality and quantities of data at desired cadence, limited only by the photon collecting power of small apertures. The data already gathered from such instruments (e.g. ELODIE, CORALIE, SOPHIE, CHIRON) are an important part of our extant canon of exoplanetary systems, and there is a growing necessity for more such workhorse RV instruments. Despite the recognition of this necessity, PARAS remains one of the few dedicated high-precision instruments of its caliber currently available to the community \cite{Fischer:2016}.

\section{Current Performance}
\label{sec:perform}

\begin{figure}
   \begin{center}
   \begin{tabular}{c}
   \includegraphics[width=0.5\textwidth]{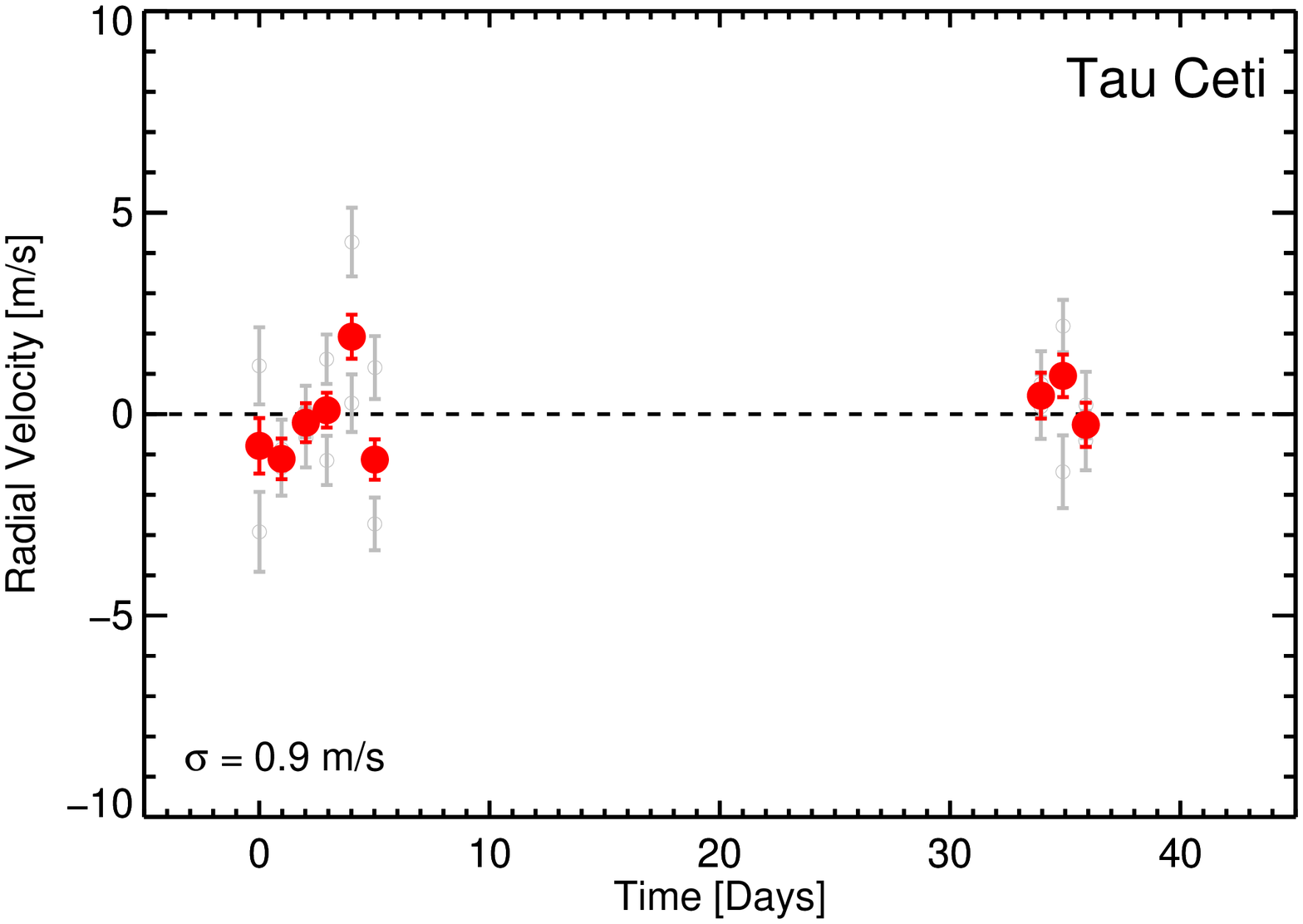}
   \includegraphics[width=0.48\textwidth]{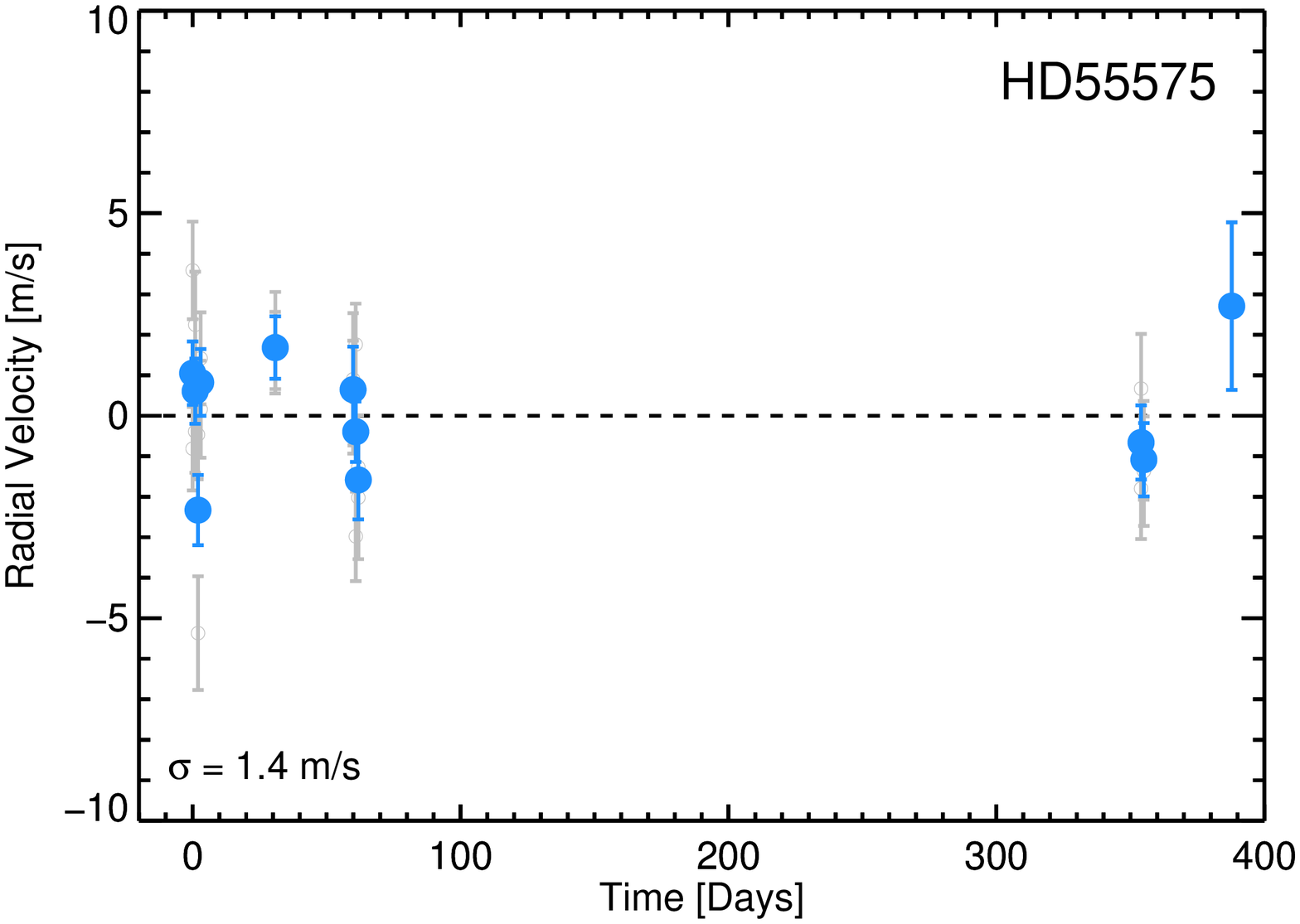}
   \end{tabular}
   \end{center}
   \caption[tauceti] 
   { \label{fig:tauceti} 
Radial velocity measurements as a function of time obtained with PARAS on bright RV standard stars. Open grey circles show individual observations while colored circles are binned nightly. {\it (Left:)} PARAS is able to achieve precisions of $<$~1~m~s$^{-1}$ over $\sim$1 month on Tau Ceti (V$_{\rm mag}=3.5$) , and {\it (Right:)} precisions of $<$~2~m~s$^{-1}$ over $\sim$1 year on HD55575 (V$_{\rm mag}=5.58$). This makes PARAS optimally poised for follow-up of bright K2 and TESS targets, as well as detailed stellar characterization and activity monitoring. }
   \end{figure}

PARAS (PRL Advanced Radial-velocity Abu-sky Search) is a high-precision spectrograph on the 1.2 m telescope in Mt. Abu, India. Owned and operated by the Physical Research Laboratory (PRL), Ahmedabad, India, PARAS was conceived as a facility-class instrument for exoplanet discovery and characterization, with 80-100 nights a year on the telescope. It was designed based on the precision velocimetry lessons learned from its predecessors, particularly HARPS, and is a stabilized fiber-fed high-resolution (R$\sim$67,000) spectrograph with broad wavelength coverage of 3800 - 9600 \AA, and simultaneous ThAr calibration \cite{Chakraborty:2014,Chakraborty:2010}. The primary goal was to target exoplanets around bright G and K dwarfs (V$_{\rm mag}<7$) with long-term RV precisions of 1-2~m~s$^{-1}$, degrading up to 5-10~m~s$^{-1}$ on fainter targets. PARAS is already successfully performing at this level (Fig.~\ref{fig:tauceti}), with further improvements anticipated in temperature control (\S\ref{sec:temp}) and telescope performance (\S\ref{sec:upgrades}). With {\bf demonstrated precisions of $<$~1~m~s$^{-1}$ over $\sim$1 month, and $<$~2~m~s$^{-1}$ over $\sim$1 year} on the RV standard stars Tau Ceti and HD55575, respectively, PARAS is now primed to lend major support to the RV follow-up effort.

\section{Lessons for future Extreme Precision Instruments}
\label{sec:lessons}

The next generation of planet hunting spectrographs will be the technological progeny of instruments like HARPS and PARAS. We focus here on two such instruments, the Habitable Zone Planet Finder (HPF) and NEID (NN-Explore Investigations with Doppler spectroscopy), that will push our ground-based capabilities in the NIR and optical, respectively. Both instruments directly inherit design choices and operational experience from PARAS, given the overlap in science and instrument teams. \\

\noindent{\bf The Habitable Zone Planet Finder} \\
\indent HPF is a stabilized fiber-fed high-resolution (R$\sim$50,000) near-infrared (NIR) spectrograph designed to detect habitable zone (HZ) planets around M-dwarfs. Currently under construction at Penn State, it will be deployed on the 10 m Hobby-Eberly Telescope (HET) in 2017 \cite{Mahadevan:2014,Mahadevan:2015}. Being a NIR instrument, HPF demands its own R\&D work for issues that are exacerbated at longer wavelengths \cite{Halverson:2015,Hearty:2014,Roy:2014a,Halverson:2014}, but still benefits directly from the lessons offered by PARAS. Both instruments have asymmetric white pupil designs, with R4 echelle gratings. The HPF RV analysis software will be based on the current PARAS pipeline (\S\ref{sec:pipe}). Most interestingly, though, the instruments have some overlap in wavelength: HPF covers 0.8 - 1.3 $ \mu$m, and PARAS extends to $\sim$9600 \AA. This allows us to test NIR calibration techniques for HPF on PARAS (\S\ref{sec:une}), and opens up interesting possibilities for complementary follow-up both chromatically and temporally, since the PRL Mt. Abu Observatory and the MacDonald Observatory are $\sim$12 hours apart. \\

\noindent{\bf NEID} \\
\indent In response to the recommendations of the Astro2010 Decadal Survey, the NASA-NSF Exoplanet Observational Research (NN-EXPLORE) partnership solicited a call for an Extreme Precision Doppler Spectrometer for the 3.5 m WIYN telescope on Kitt Peak. The selected instrument,
NEID, is a high-resolution (R$\sim$100,000), ultra-stable optical spectrometer being built by a multi-institutional and interdisciplinary team led by Penn State. Aiming at RV precisions $<$30~cm~s$^{-1}$, it represents the first wave of extreme precision doppler instruments capable of detecting true Earth twins and identifying targets for future space missions such as WFIRST-AFTA, HabEx, or LUVOIR. NEID is designed as a complete radial velocity system, incorporating ultra-stable environmental control, cutting edge wavelength calibration, innovative fiber feed design, and broad wavelength coverage with a suite of chromospheric activity indicators. 
 
 In many ways, NEID is the true successor of PARAS, with a classical white pupil design, and large monolithic prism cross-disperser that provides a wide 3800-9300 \AA~bandpass on a single detector at high efficiency. Both prisms are made of i-line Ohara glasses that have high homogeneity and excellent transmission into the blue (PBM2Y for NEID, and PBM8Y for PARAS), and positive experience with PARAS was an integral part of this design choice. As a baseline, NEID inherits the PARAS data pipeline (\S\ref{sec:pipe}), although it will be refined for performance at the 30~cm~s$^{-1}$ level with a laser frequency comb calibrator. NEID also uses a deep depletion e2v CCD, very similar to the PARAS detector, which allows us to begin preliminary characterization on a functioning science-grade CCD, and to develop methods to quantify and mitigate issues like fringing and charge diffusion caused by real astrophysical observations (\S\ref{sec:ccd}). The NEID baseline input optics design borrows heavily from PARAS, as does the overall optical alignment plan. Ongoing PARAS operations and an accumulating data stream have also proved valuable in many other ways, including selection of bright, quiet targets with long baselines of observation, and the NEID operations concept for calibration and maintenance. Additionally, there is the interesting possibility of NEID and PARAS working in tandem to provide complementary follow-up given that the observatories are 12.3 hours apart.
  
   \begin{figure}[ht]
   \begin{center}
   \begin{tabular}{c}
   \includegraphics[width=0.65\textwidth]{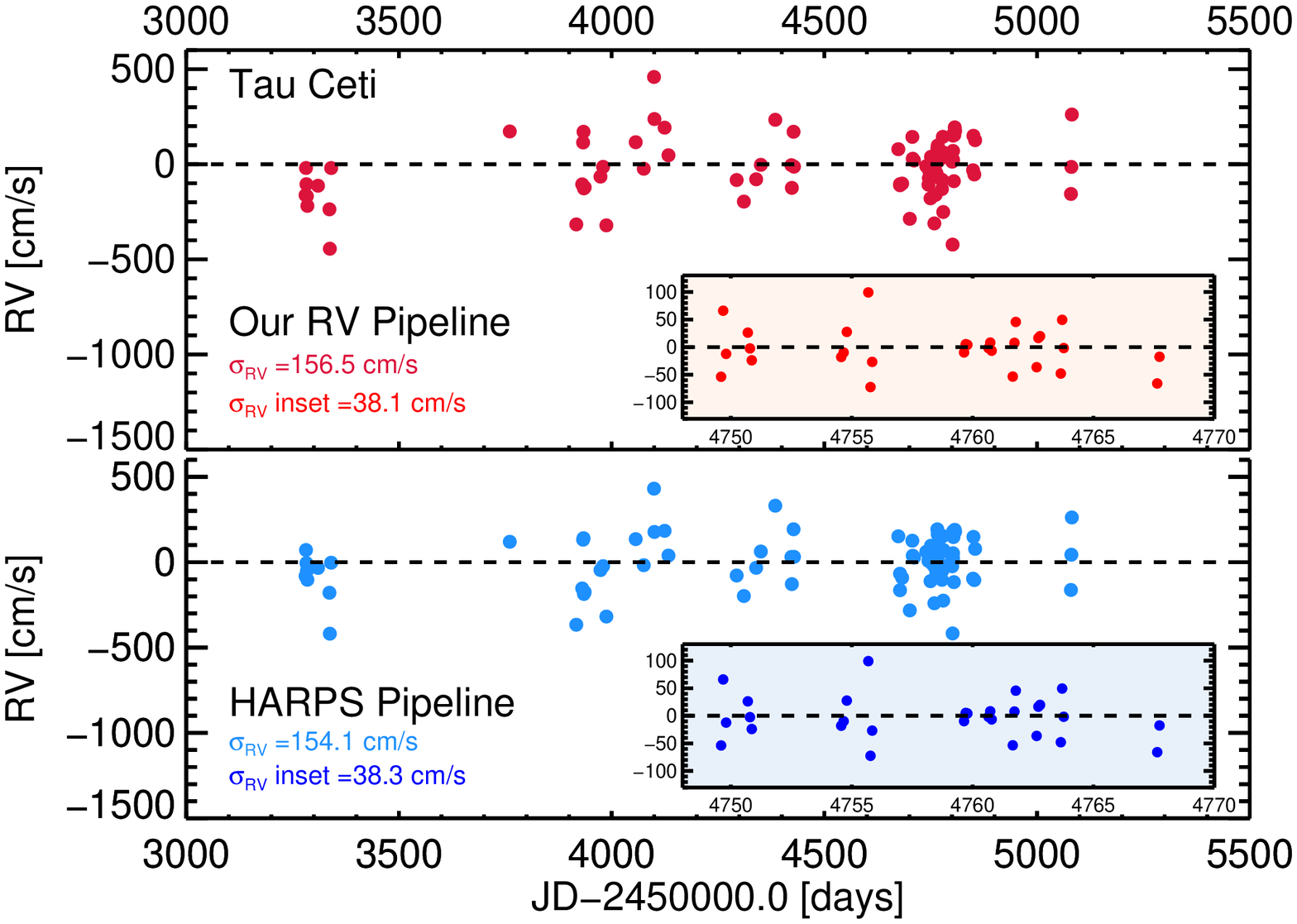}
         \includegraphics[width=0.3\textwidth]{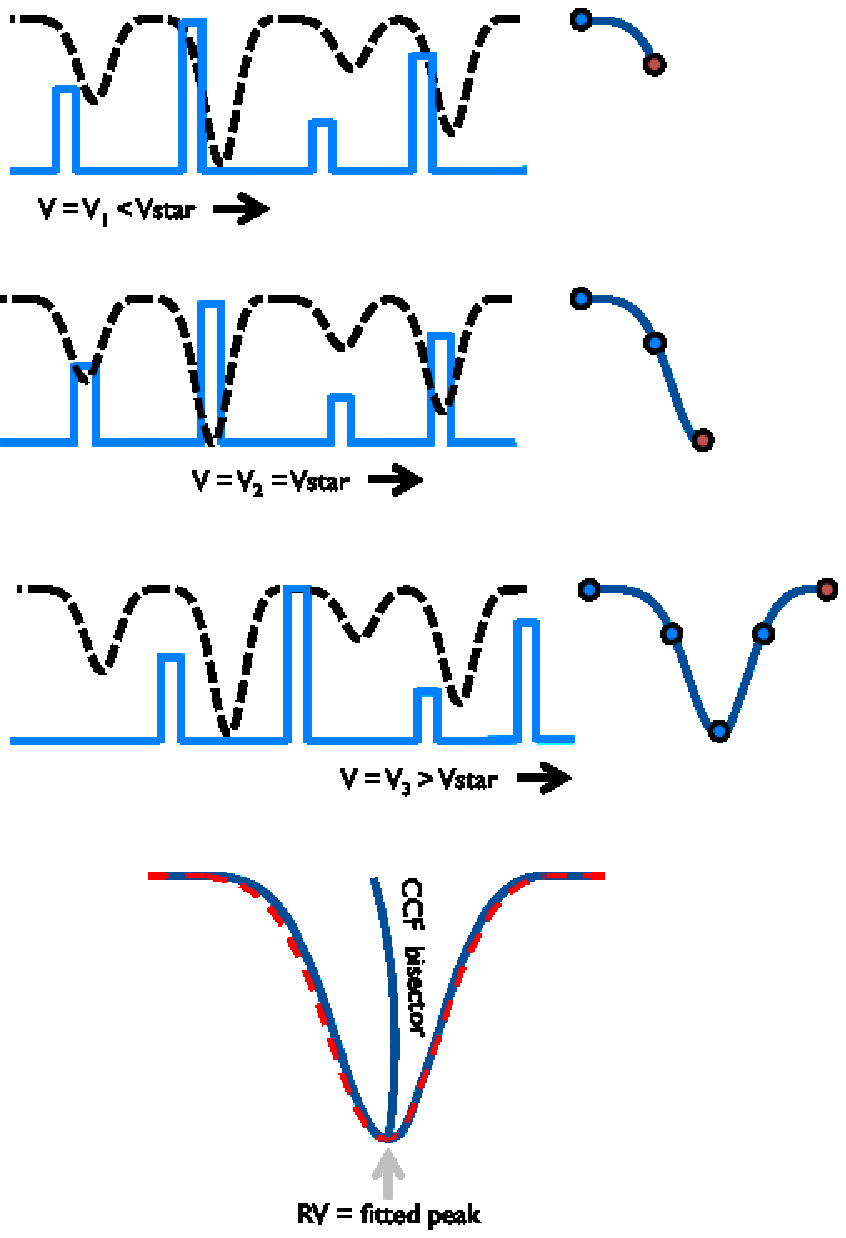}
   \end{tabular}
   \end{center}
   \caption[pipe] 
   { \label{fig:pipe} 
As a baseline, NEID inherits the PARAS data reduction pipeline. The PARAS pipeline currently performs at the same level as the public output of the HARPS pipeline. {\it (Left:)} We verified the technical capability of the PARAS RV analysis algorithms by applying it to 84 HARPS observations of the standard star Tau Ceti, taken over a five year span. Insets show a subset of data collected with an optimized HARPS observing strategy \protect\cite{Dumusque:2011, Pepe:2011} with three observations per night, and long integrations to average over stellar p-modes. {\it (Right:)} Schematic of mask based cross-correlation for CCF and RV recovery \protect\cite{Baranne:1996,Pepe:2002}.}
   \end{figure} 
   
\subsection{Precision RV Pipeline}
\label{sec:pipe}
In the era of extreme precision RV, every aspect of instrumentation, observation, and analysis come under scrutiny and must be refined. Groundbreaking and believable exoplanet discovery thus relies heavily on the data analysis pipeline. The automated PARAS data reduction pipeline (DRP), described in detail in Chakraborty et al. (2014), performs image processing, optimal extraction, simultaneous ThAr wavelength calibration, and then determines RV measurements via mask cross-correlation \cite{Baranne:1996,Pepe:2002}, with corrections for barycentric motion and instrument drift. Although it already performs at the same level as the HARPS public pipeline (Fig.~\ref{fig:pipe}), it is regularly upgraded to meet the needs of the instrument. As an illustrative example, the prism thermal response to temperature variations on the optical bench (\S\ref{sec:temp}) was causing variations in cross dispersion and hence order location on the detector. The slightly erroneous extractions with the early pipeline had an insidious effect on the RVs. Although we are expecting much better temperature control in the future, the pipeline was consequently equipped to respond to both shifts and stretches in the order re-trace. As PARAS moves from instrument characterization to survey mode, the pipeline is being evolved to meet observer needs, with new modules for faint stars, M dwarfs, and UNe calibration. Version control is also being imposed more strictly as the use of the software propagates beyond the core instrument team.

The PARAS DRP will form the basis for the HPF and NEID RV analysis software. Algorithmic precision must be maximized in response to the scant overall error budgets for these next-generation instruments (Halverson et al., {\it in these proceedings}), and there are certain conspicuous areas for improvement. For example (a) moving beyond current optimal extraction routines \cite{Piskunov:2002} to more flexible low-noise methods, (b) applying high precision barycentric correction with measurement of the wavelength dependent flux-weighted mid-exposure time  \cite{Wright:2014}, and (c) utilizing the full potential of the laser frequency comb calibrator for individual pixel characterization and the careful treatment of CCD stitch boundaries \cite{Bauer:2015}. Essentially, every subroutine will need to be optimized, and new RV extraction and stellar activity removal methods tested. However, the exploration of these methods will be grounded firmly in the proven pipelines of the previous generation of precision RV instruments.

\subsection{CCD Fringing}
\label{sec:ccd}

   \begin{figure}
   \begin{center}
   \begin{tabular}{c}
   \includegraphics[width=0.6\textwidth]{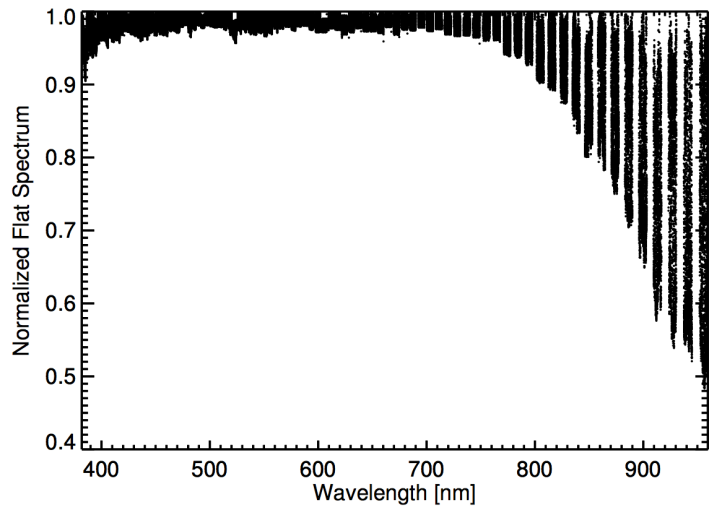}
   \end{tabular}
   \end{center}
   \caption[famp] 
   { \label{fig:famp} 
    An extracted flat frame shows the order by order wavelength dependent fringe spectrum in a the PARAS deep depletion e2v CCD231-84-1-D81. In the reddest orders, the amplitude is $\sim$50\% of the normalized flux. Gaps between orders are due to the deliberate truncation of low-signal edges that are not properly deblazed.}
   \end{figure} 

Fringing results from interference between incident and reflected beams within the CCD layers, becoming more pernicious at longer wavelengths where the decreasing absorption coefficient for silicon makes a larger proportion of the photons available for interference, as opposed to absorption during first passage. Spectroscopic observations are prone to fringing due to the delivery of essentially monochromatic light to each resolution element, in contrast to imaging where bright sky emission lines are the primary culprit. Furthermore, the structure of the AR coating, varying incident angles across the beam, and the fact that the front and back surfaces of a CCD are not exactly parallel or planar, result in irregularly shaped fringes \cite{Malumuth:2003}.

PARAS uses a 4kx4k deep depletion e2v CCD231-84-1-D81 with a custom ASTRO-broadband antireflection (AR) coating for high quantum efficiency across the wavelength range \cite{Chakraborty:2014}. Despite the 40 $\mu$m thickness of the substrate, etaloning creates noticeable fringing in the reddest orders (Fig.~\ref{fig:famp}). NEID will use a very similar 9kx9k e2v CCD with an ASTRO Multi-2 coating, and PARAS provides the rare opportunity to formulate a post-processing fringing cure for NEID with on-sky data. As an aside, our extensive experience with the PARAS deep depletion detector suggests that charge diffusion, which can result in a PSF that is flux dependent, should not be a major issue for NEID, even at 400 nm. 

\begin{figure}
   \begin{center}
   \begin{tabular}{c}
   \hspace*{-0.4in}
   \includegraphics[width=0.55\textwidth]{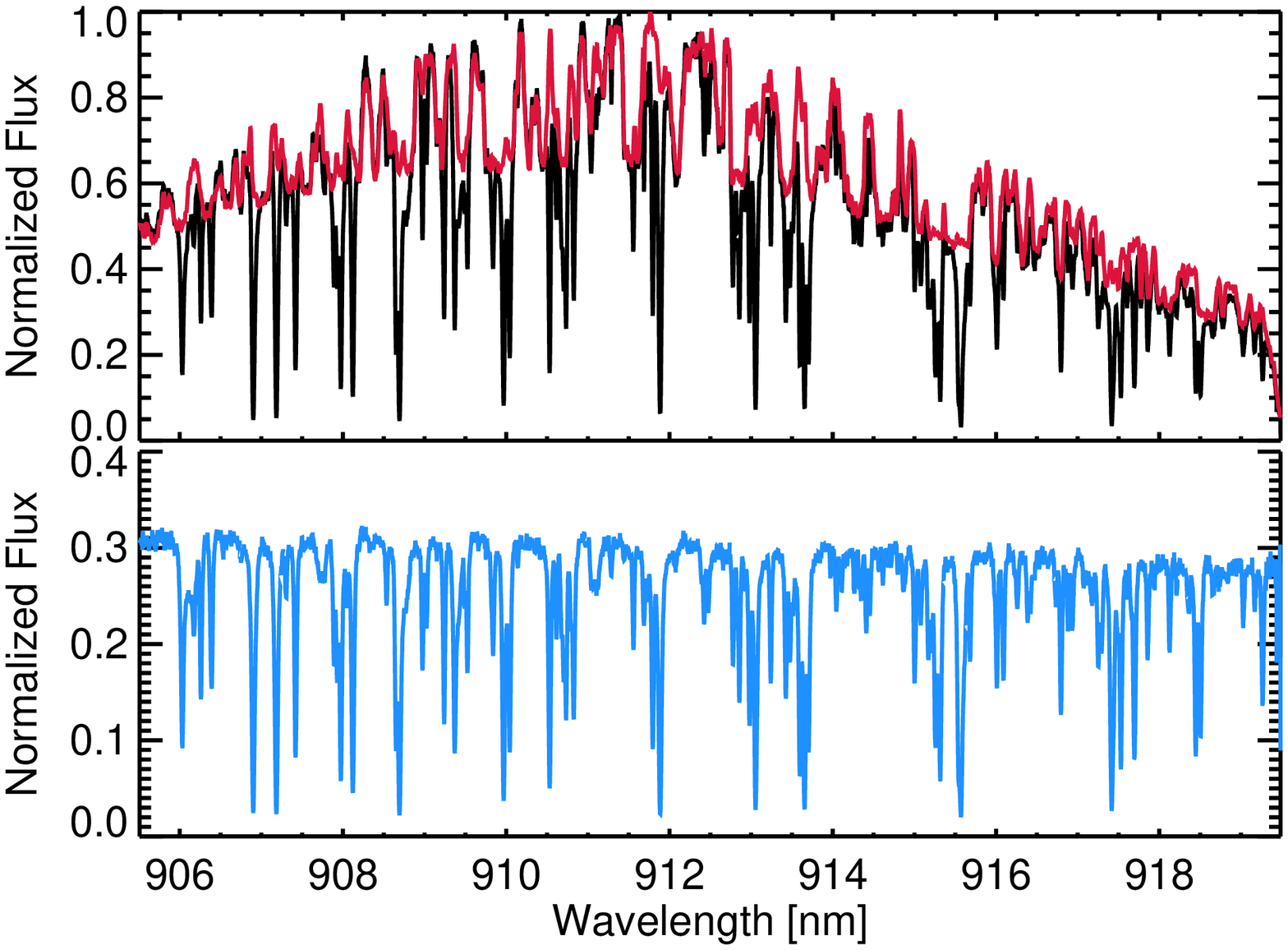}
      \hspace*{-0.2in}
      \includegraphics[width=0.53\textwidth]{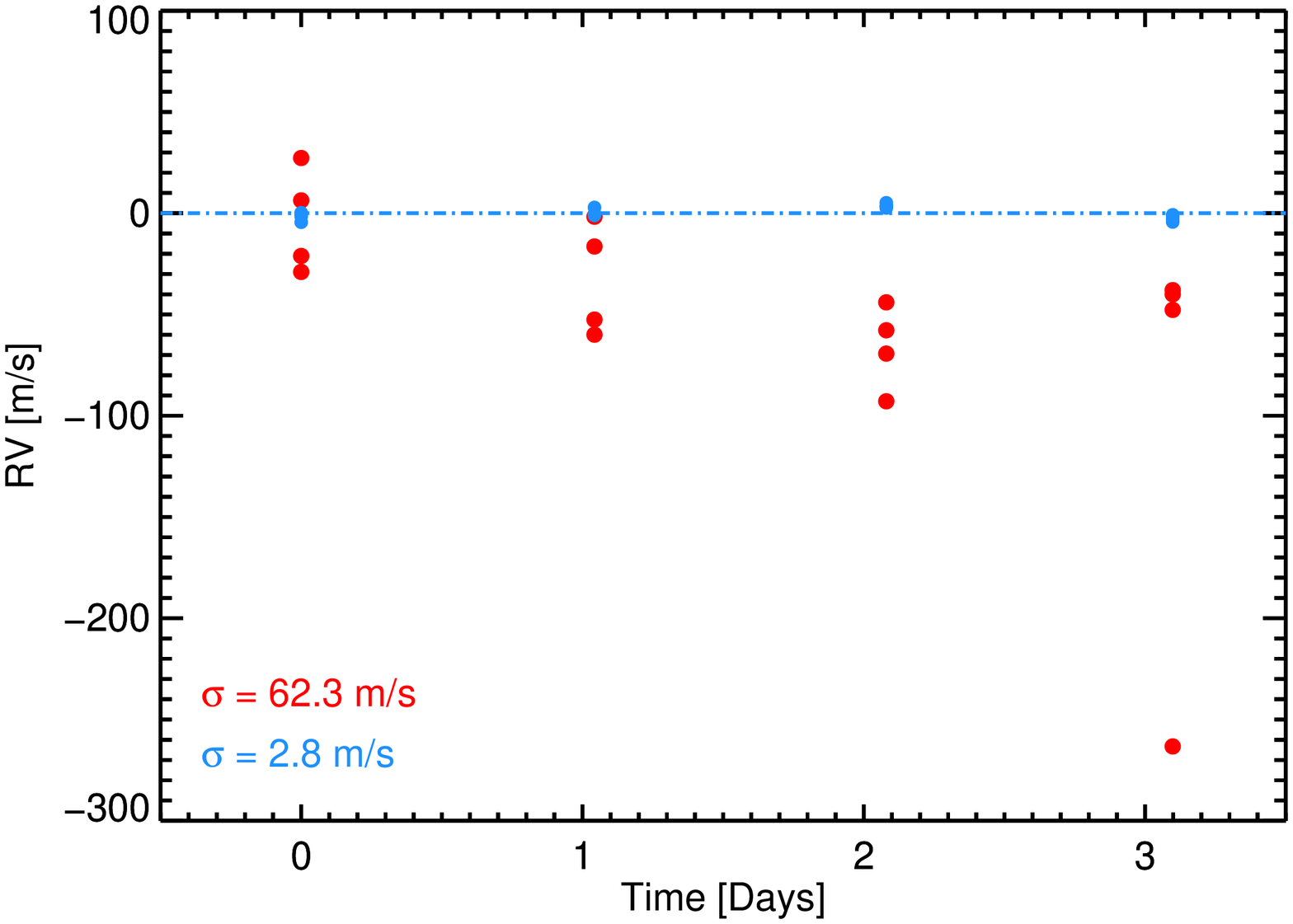}
   \end{tabular}
   \end{center}
   \caption[fringe] 
   { \label{fig:fringe} 
    Fringing is an issue in the reddest PARAS orders, despite the 40 $\mu$m thickness of the deep depletion e2v CCD. {\it (Top Left:)} One echelle order of the observed spectrum of Vega (black), and an extracted contemporaneous flat frame (red). {\it (Bottom Left:)} Vega, with the extracted flat order divided out. To first order, the flat provides good fringe correction. {\it (Right:)} The effect of uncorrected fringing on RV extraction. Variations in fringing degrade RV precision in a subset of red orders (order 86-92, 8300 - 9050 \AA) compared to precision achieved for a subset of blue orders (order 42-47, 5150 - 5450 \AA). Simply dividing out a single flat improves the red RV precision to $\sigma$=24.4 m s$^{-1}$.}
   \end{figure}

There are certain fringe suppression technologies offered by e2v, including `roughening' of the sensor back surface, or the use of a graded AR coating, but these are difficult and expensive to execute on large-format CCDs like the one required for NEID. Furthermore, these are unproven technologies in terms of precision RV, and contribute risk to the aggressive NEID budget and schedule. More typically, contemporaneous flats are used to correct fringing in the reduction pipeline, but the residual effect on RVs is still not well quantified. Figure \ref{fig:fringe} {\it(left)} shows the spectrum of Vega overlaid with an extracted flat frame (the fringe spectrum as produced by a sum-extraction), and the results from simply dividing out the flat spectrum. Although this looks promising, we quantify the effect on RVs using a synthetic M2 BT Settl stellar spectrum (T$_{\rm eff}$ = 3500 K). The synthetic spectrum is modulated by 16 flat frames taken with PARAS on 4 consecutive days, at the beginning of each night. This is then processed with the PARAS RV analysis code, using a custom numerical M2 stellar mask. 

Figure \ref{fig:fringe} {\it(right)} shows the effect of uncorrected fringing on RV extraction. The variations in fringing, within and between nights, significantly degrade RV precision in a subset of red orders (order 86-92, 8300 - 9050 \AA) compared to precision achieved for a subset of blue orders (order 42-47, 5150 - 5450 \AA). Note that the RV precision in the blue is limited by the fact that we are fitting a Gaussian to a markedly non-Gaussian cross-correlation function (CCF) for these M-dwarfs, where the lack of a clear continuum can produce a lot of structure in the wings. Dividing out a single representative flat improves the red RV precision to $\sigma$=24.4 m s$^{-1}$, indicating that variations in fringing, even within a night, need careful treatment beyond dividing out beginning of the night calibration flats. The NEID ASTRO Multi-2 coating is an improvement on the PARAS coating, and will mitigate the fringing to some degree, However, this is a continuing effort to ensure that we can maximize the utility of the red wavelength coverage in NEID and PARAS, for both stellar activity indicators and RV information content.        

\subsection{Temperature Control}
\label{sec:temp}
\begin{figure}
   \begin{center}
   \begin{tabular}{c}
   \includegraphics[width=0.6\textwidth]{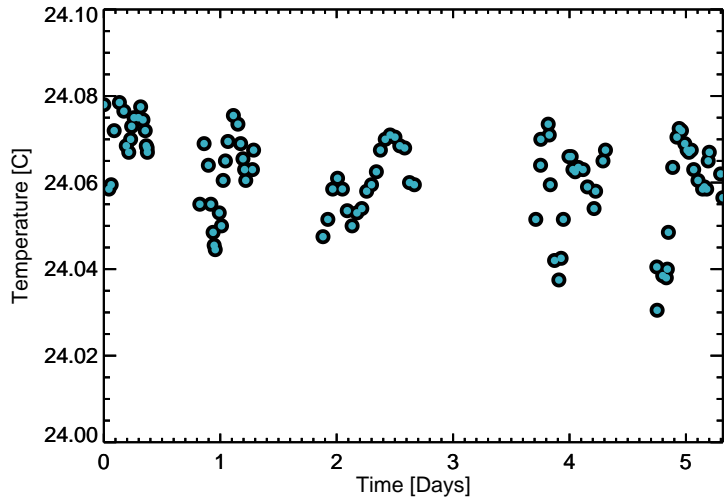}
   \end{tabular}
   \end{center}
   \caption[temp] 
   { \label{fig:temp} 
    The improved PARAS temperature control system uses a sensor on the optical bench, close to the base of the prism, to adjust additional heaters. We are currently able to control at 24.06$\pm$0.01$^{\circ}$C, with the expectation that this will improve to $\pm$0.005$^{\circ}$C in the near future with the full implementation of the new system.}
   \end{figure}

In spite of careful environmental control design \cite{Chakraborty:2014}, PARAS RVs showed seasonal trends in early data. The instrument lives inside two concentric insulation rooms, with temperature control systems in both. The vacuum vessel sits on eight trisolator legs on an insulated and vibration-isolated concrete pier. However, notwithstanding these measures there is heat leakage between the base of the vessel and the pier, leading to imperfect thermal decoupling of the instrument from the mountain. The implementation of additional heaters, air circulation fans, and extra calibrated heat pads, as well as a new temperature sensor on the optical bench has largely solved this problem. The extra heaters are controlled by a custom electronics system for differential response to the sensor, placed close to prism since it is the most thermally vulnerable optical component. Prior to the new system, there was clear wavelength dependent variation in the order locations, and blue orders were shifting more than the red orders. Using the specifications provided by Ohara for the PBM8Y glass (dn/d$\lambda$, dn/dT), we were able to match the order shifts very closely to the modeled prism thermal response, indicating temperature variations up to 0.1$^{\circ}$C. Currently we have improved temperature control at the level of 24.06$\pm$0.01$^{\circ}$C (Fig.~\ref{fig:temp}). With the new system in full automated operation, we expect to control at 24$\pm$0.005$^{\circ}$C. Although these thermal issues are inherent to the underlying concrete pier structure and hence unavoidable for PARAS, they bequeath an important lesson to the environmental control systems of our new instruments, HPF and NEID.

\subsection{UNe Calibration Testbed}
\label{sec:une}
\begin{figure}
   \begin{center}
   \begin{tabular}{c}
   \includegraphics[width=0.6\textwidth]{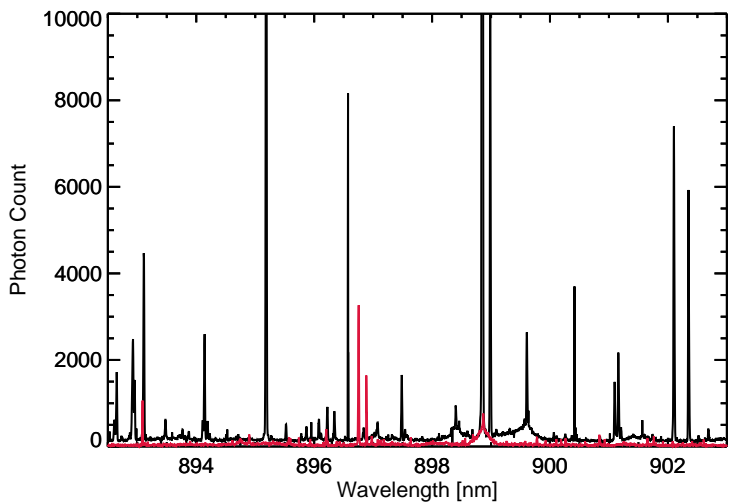}
   \end{tabular}
   \end{center}
   \caption[une] 
   { \label{fig:une} 
    PARAS observations showing UNe (black) and ThAr (red) calibration spectra. UNe has stronger lines and a richer line density in the NIR, without the contaminating effects of argon\protect\cite{Redman:2011}. Preliminary results indicate that PARAS can successfully operate in a bracketed UNe calibration mode to observe bright early M dwarfs.}
   \end{figure}

Although ThAr is commonly used for wavelength calibration in the optical (3800 - 6800 \AA) \cite{Redman:2014,Lovis:2007}, we attempt to take full advantage of the red coverage of PARAS using a UNe lamp, which is better suited for the red-optical and NIR (Fig.~\ref{fig:une}) \cite{Redman:2011}. This is a new mode of operation for PARAS, and we have begun observations of certain bright M dwarfs (e.g. GJ411, HD79210) to gauge the precision achievable with bracketed UNe calibration for these targets.  Preliminary results show good agreement with the Redman et al. (2011)\cite{Redman:2011} and Palmer et al. (1980)\cite{Palmer:1980} Fourier Transform Spectrograph (FTS) line lists. The wavelength overlap between PARAS and HPF proves highly beneficial in this case, since it allows us to develop a new list of uranium and neon emission lines from high-resolution spectroscopy (Roy et al. {\it in prep}), in anticipation of HPF calibration requirements. Despite fringing issues, the ability to calibrate the redder part of the spectrum with UNe opens up PARAS' capability for observing early M-dwarfs and provides a testbed for NEID's performance on cool stars.

\section{Scheduled Upgrades for PARAS}
\label{sec:upgrades}

Several improvements have already been implemented on PARAS, in concert with instrument characterization. These include enhanced scrambling with the addition of a 50 $\mu$m core octagonal fiber section, regular upgrades to the data reduction pipeline, and added elements for temperature control. The full implementation of the new temperature control system is imminent, which will allow PARAS to reliably gather long baselines of data. However, PARAS is still limited by the performance and size of the 1.2 m telescope. The telescope system is aged and suffers from hysteresis in the RA/Dec drives, flexure issues, and has tracking accuracies of only 1-1.5". To counter these issues, the fiber is undersized with respect to the seeing (1.8" on sky, while seeing varies between 1.3-2.0"). While this lends PSF stability in median seeing conditions, it also leads to light loss. In addition, the reflectivity of the 1.2 m telescope primary can rapidly degrade in bad weather (50-60\%), despite attempts at regular wet cleaning and re-aluminization. In order to optimize scientific yield, a new state-of-the-art robotic 2.5~m telescope is under construction, next to the current facility. PARAS will move to the new telescope in 2019, and receive up to 50\% of the telescope time for exoplanet science. With high-precision RV capability, coverage of a suite of stellar activity indicators, and a substantial guaranteed time allotment on the new telescope, we thus present PARAS as a true workhorse RV instrument for the exoplanet community.

\acknowledgments     
 
Funding for the PARAS project is provided by the Department of Space, Government of India through the Physical Research Laboratory. We thank the PRL Director for supporting the PARAS program. This work was partially supported by funding from the Center for Exoplanets and Habitable Worlds. The Center for Exoplanets and Habitable Worlds is supported by the Pennsylvania State University, and the Eberly College of Science. We acknowledge support from NSF grants AST 1006676, AST 1126413, AST 1310885, and the NASA Astrobiology Institute (NNA09DA76A) in our pursuit of precision radial velocities in the optical and NIR. AR acknowledges support from the Penn State Downsbrough Graduate Fellowship Program and the Lewis \& Clark Fund for Exploration and Field Research in Astrobiology. AC is grateful to Larry Ramsey (Pennsylvania State University) and Francesco Pepe (Geneva Observatory) for detailed discussions over the years on aspects of building precision RV spectrographs. This work has made use of the SIMBAD database (operated at CDS, Strasbourg, France), and NASA's Astrophysics Data System.

\bibliography{simple}   
\bibliographystyle{spiebib}   

\end{document}